\documentclass[11pt]{article}
\usepackage{epsfig}
\newcommand{\sss}{\vspace{.2in}}

\newcommand{\be}{\begin{equation}}
\newcommand{\ee}{\end{equation}}
\newcommand{\bea}{\begin{eqnarray}}
\newcommand{\eea}{\end{eqnarray}}
\newcommand{\sn}{{\rm sn}}
\newcommand{\cn}{{\rm cn}}
\newcommand{\dn}{{\rm dn}}
\newcommand{\cs}{{\rm cs}}

\begin{document}
~\hfill{\footnotesize SUNYB/04-03, IP/BBSR/04-05,~~\today}
\sss
\sss
\begin{center}
{\Large {\Large \bf Periodic Potentials and Supersymmetry}}
\end{center}
\vspace{.5in}
\begin{center}
{\large{\bf
   \mbox{Avinash Khare}$^{a,}$\footnote{khare@iopb.res.in} and
   \mbox{Uday Sukhatme}$^{b,}$\footnote{sukhatme@buffalo.edu}
 }}
\end{center}
\vspace{.6in}
\noindent
a) \hspace*{.2in}
Institute of Physics, Sachivalaya Marg, Bhubaneswar 751005, Orissa, India\\
b) \hspace*{.2in}
Department of Physics, State University of New York at Buffalo,
Buffalo, NY 14260, U.S.A. \\
\sss
\sss
\begin{abstract}
We review the current status of one dimensional periodic potentials and also present several
new results. It is shown that using the
formalism of supersymmetric quantum mechanics, one can considerably enlarge the limited class of
analytically solvable one-dimensional periodic potentials. Further,
using the Landen transformations as well as cyclic identities for Jacobi elliptic functions 
discovered by us recently, it is shown that a linear superposition of
Lam\'e (as well as associated Lam\'e) potentials are also analytically
solvable. Finally, using anti-isospectral transformations, we also
obtain a class of analytically solvable, complex, PT-invariant,
periodic potentials having real band spectra.  
\end{abstract}
\newpage

\sss
{\noindent\bf I. INTRODUCTION} 

One dimensional potential wells have bound states. They are solutions
of the Schr\"odinger equation which satisfy appropriate boundary
conditions. When two identical potential wells are very far apart,
then each potential has the same energy levels, and each eigenstate is
doubly degenerate. As the wells are brought closer together, there is
communication between them, and each level is split into two.
Similarly, if one has an array of many identical wells making a
periodic potential, then one gets energy bands, which play, for
example, a crucial role in determining the electronic properties of 
crystalline solids. To illustrate this band structure qualitatively,
condensed matter physics texts usually treat the problem of a one
dimensional periodic array of delta functions, called the Kronig
Penney model. Here, one gets a transcendental equation for computing
band edges. Another well-studied class of periodic potentials are the
Lam\'e potentials \cite{mw,ar,ww}
\be\label{1}
V(x)=a(a+1) m \sn^{2} (x,m)\,, 
\ee
where $\sn(x,m)$ is a Jacobi elliptic function \cite{gr} of real elliptic
modulus parameter m ($0 \le m \le 1$). If $a$ is any positive integer,
it is well known that these potentials have only $a$
band gaps and hence $2a+1$ band edges which are analytically 
known (in principle). 
However, this is an exceptional example. The bottom line
is that even in one dimension, there are very few solvable periodic
potentials, and it would be desirable to have more, especially some with a
richer spatial structure.

We shall describe how to obtain new solvable periodic potentials via
several different approaches. One way, for example is to expand our
knowledge of Lam\'e potentials to the wider class of associated Lam\'e  
(AL) potentials \cite{ks1,ks2}
\be\label{2}
V(x)=a(a+1) m\sn^2(x,m)+b(b+1) m\frac{\cn^2(x,m)}{\dn^2(x,m)}~,
\ee
where $a,b$ are constants and without any loss of generality 
we consider $a \ge b$. 
A second way, for example is to exploit the fact that Jacobi elliptic
functions are doubly periodic functions and 
consider PT-invariant complex periodic potentials \cite{bed}
obtained from the Lam\'e and the AL potentials 
by considering 
the anti-isospectral transformation
$x \rightarrow ix+\beta$ (where $\beta$ is any nonzero real number)
\cite{ks3}.
A third way is to consider a linear superposition of several Lam\'e 
(AL) potentials which in view of the Landen transformations
\cite{lau,ks4}
and cyclic identities \cite{ks5} for Jacobi elliptic functions can be shown to be 
essentially equivalent to 
Lam\'e (AL) potentials with modified
parameters.
Finally, we can further expand all these classes of solvable
potentials by using the techniques of supersymmetric quantum mechanics to generate 
supersymmetric partner potentials \cite{cks}.   

The outline of this paper is as follows. In Sec. II we review some
general properties of periodic potentials as well as
the relevant formalism of supersymmetric
quantum mechanics. In Sec. III we discuss the band edge energies and
eigenfunctions for several Lam\'e and AL potentials. In particular, for
Lam\'e potentials (\ref{1}) we derive a
remarkable hitherto un-noticed relation connecting band edge energy eigenstates corresponding to   
modulus parameters $m$ and $1-m$. 
We also 
describe how the Lam\'{e} potential results can be vastly expanded to get 
solutions for the AL potentials (\ref{2}). In
particular, we show that for any integral $a,b$ with 
$a > b$, there are  $a$ bound bands  
followed by a continuum band, out of which the top $b$ bound bands are
unusual in the sense that both band edges are of period $2K(m)$
($4K(m$)) when the integer $a-b$ is even(odd), 
where $K(m)$ is the complete elliptic integral of the first kind.
In Sec. IV we obtain
the supersymmetric partners of both the Lam\'{e} and the AL 
potentials, thereby expanding the list of solvable periodic potentials. 
In Sec. V we show that using Landen transformations \cite{ks4} and
cyclic identities obtained by us recently \cite{ks5}, exact
band edge eigenvalues and eigenfunctions can also be obtained in the
case of linear
superposition of Lam\'e and AL potentials. In Sec. VI
we show that a class of PT-invariant complex periodic potentials obtained from
the Lam\'e and AL
potentials by the anti-isospectral transformation $x \rightarrow ix+\beta$,
($\beta$ being any nonzero real number) are also exactly solvable
problems with real band spectra. Finally, in Sec. VII we discuss 
the double sine-Gordon equation and 
show that in some special cases it has  
unusual band spectra, in that the majority of bound bands have both of
their band edges of the same period. 

\vspace{.2in}
{\noindent\bf II. GENERAL PROPERTIES} 

\noindent{\bf (a) Periodic Potentials:} For a real potential with period $L$, 
one is seeking solutions of the Schr\"{o}dinger equation subject to the 
Bloch condition
\be\label{2.1}
\psi (x) = e ^{ikL} \ \psi (x+L) \, ,
\ee
where $k$ denotes the crystal momentum. The spectrum shows energy bands whose 
edges correspond to $kL=0,\pi$, that is the wave functions at the band 
edges satisfy $\psi (x) = \pm \psi (x+L)$. For periodic potentials, the 
band edge energies and wave functions are often called eigenvalues and 
eigenfunctions, and we will also use this terminology.

A general property of eigenstates for a real potential with period $L$ is the 
oscillation theorem. It states that band edge wave functions, when arranged 
in order of increasing energy 
$E_0 < E_1 \le E_2 < E_3 \le E_4 < E_5 \le E_6 < ...$,
have periods $L,2L,2L,L,L,2L,2L,...$ \cite{mw}. 
The corresponding number of wave function nodes in the interval 
$L$ are $0,1,1,2,2,3,3,...$ and the energy band gaps are given by 
$\Delta_1 \equiv E_2 - E_1,~\Delta_2 \equiv E_4 - E_3,~ 
\Delta_3 \equiv E_6 - E_5,~... $\,. The oscillation theorem 
is very useful in identifying if all band edge eigenstates have been 
properly determined or if some have been missed. 

{\noindent\bf (b) Supersymmetric Quantum Mechanics:}
The supersymmetric partner potentials $V_{\pm}(x)$
are defined in terms of the superpotential $W(x)$ by \cite{cks}
\be\label{2.2}
V_{\pm} (x) = W^2 (x) \pm W'(x) \, ,
\ee
where (by convention) the superpotential $W(x)$ is related to the
ground state eigenfunction of $V_{-} (x)$ by
\be\label{2.2a}
W(x) = -\frac{\psi_{0}'^{(-)}(x)}{\psi_{0}^{(-)} (x)}\,.
\ee
The corresponding Hamiltonians $H_{\pm}$ can be factorized as
\be\label{2.3}
H_{-} = A^+ A,~~  H_{+} = A A^+ \, ,
\ee
where
\be\label{2.4}
A = {d\over dx} + W(x) \, , ~~A^+ = -{d\over dx} + W(x) \, ,
\ee
so that the spectra of $H_{\pm}$ are nonnegative. It is also
clear that
on the full line, both $H_{\pm}$ cannot have zero energy modes, since
the two functions $\psi_0^{(\pm)}$ given by
\be\label{2.5}
\psi_0^{(\pm)} (x) = \exp (\pm \int^x W(y) dy) ~~ ,
\ee
cannot be simultaneously normalized.

On the other hand, when the superpotential
$W(x)$ has 
period $L$ [$W(x+L)=W(x)$], and the eigenfunctions of both
$V_{\pm}(x)$ must satisfy the Bloch condition (\ref{2.1}), it is
easily shown that in the periodic case, irrespective of whether
supersymmetry is broken or unbroken, the spectra of $V_{\pm} (x)$ are
strictly identical \cite{df,bg}. 
Further, one has unbroken supersymmetry 
provided
\be\label{2.6}
\int^L_0 W(y) dy = 0 \, ,
\ee
and in this case the partner Hamiltonians have identical spectra
including zero modes and both $\psi_0^{(\pm)}$ belong to the 
Hilbert space. 
As a result, unlike on the full line, for
periodic potentials, whenever the condition (\ref{2.6}) is satisfied,
supersymmetry is unbroken and yet the Witten index (which
counts the difference between the number of zero modes of
$\psi_0^{(\pm)}$) is always zero.   

The condition
(\ref{2.6}) is trivially satisfied when $W(x)$ is an odd function of $x$
and throughout this paper we shall only consider 
superpotentials $W$ which are odd 
function of $x$.
Further, using the
known eigenfunctions $\psi^{(-)}_n (x)$ of $V_{-}(x)$,
one can
immediately
write down the corresponding  eigenfunctions $\psi^{(+)}_n (x)$
of $V_{+} (x)$. In particular, from eq. (\ref{2.5}) it follows that the 
ground state of $V_{+} (x)$ is given by \cite{cks}
\be\label{2.7}
\psi^{(+)}_{0} (x) = {1\over \psi^{(-)}_{0} (x)} \,~ ,
\ee
while the un-normalized excited states $\psi^{(+)}_{n} (x)$ are
obtained from $\psi^{(-)}_{n} (x)$
by using the relation 
\be\label{2.8}
\psi^{(+)}_{n} (x) = \left[{d\over dx} +W(x)\right] \psi^{(-)}_{n} (x) ~,~ (n \ge 1)~.
\ee
Thus by starting from an exactly solvable periodic potential $V_-(x)$, 
one gets a new isospectral periodic
potential $V_+(x)$. 

A few years ago, the concept of self-isospectral periodic potentials 
has been defined and developed  
in detail \cite{df}. A one 
dimensional potential $V_-(x)$ of period $L$ is said to 
be self-isospectral if its supersymmetric partner 
potential $V_+(x)$ is just the original potential upto a 
discrete transformation - a translation by any constant amount, 
a reflection, or both. A common example is translation by half a 
period, in which case the condition for self-isospectrality is 
$V_+(x)=V_-(x-L/2).$
In this sense, any self-isospectral potential is rather uninteresting, since  
the application of supersymmetry just yields a discrete transformation and 
basically nothing new. 

\vspace{.2in}
{\noindent\bf III. LAM\'E AND ASSOCIATED LAM\'E POTENTIALS}

\noindent{\bf (a) Lam\'{e} Potentials:}
The Lam\'e potentials as given by eq. (\ref{1}) 
have a period $L=2K(m)$. Their name comes from the fact that the 
corresponding Schr\"{o}dinger 
equation (with $\hbar=2m=1$)
\be\label{3.1}
-\frac{d^2 \psi}{dx^2} + a(a+1)m{\rm sn}^2(x,m) \psi = E \psi~,
\ee 
is called Lam\'{e} equation \cite{mw,ar}.  
It is well known 
that for any integer value $a=1,2,3,\ldots$, the corresponding 
Lam\'{e} potential has $a$ bound bands followed by a continuum band 
\cite{mw,ar} and the $2a+1$ band edge energy eigenstates
are analytically known in principle.  
We now obtain remarkable new relations (valid for any integer $a$)
relating the band edge energy eigenvalues and eigenfunctions at two
values $m$ and $1-m$ of the modulus parameter.

We start from the Schr\"odinger equation (\ref{3.1}). On using
the relation \cite{gr}
\be\label{3.2}
\sqrt{m}\,\sn(x,m)=-\dn[ix+K'(m)+iK(m),1-m]\,,
\ee
and defining a new variable $y=ix+K'(m)+iK(m)$, the
Schr\"odinger eq. (\ref{3.1}) takes the form 
\be\label{3.3}
-\psi''(y)+a(a+1)(1-m)\sn^2(y,1-m) \psi(y) = [a(a+1)-E] \psi(y)\,.
\ee
On comparing eqs. (\ref{3.1}) and (\ref{3.3}) we then have the desired 
relations 
relating the eigenstates when the modulus parameter is  $m$ and $1-m$:
\be\label{3.4}
E_j (m) = a(a+1) -E_{2a-j} (1-m)\,,~~\psi_{j} (x,m) \propto 
\psi_{2a-j} (y,1-m)\,,~~j=0,1,...,2a\,.
\ee
Thus, for any
integer $a$, at
$m=1/2$, one has remarkable relations  
\bea\label{3.5}
&&E_j (m=1/2) +E_{2a-j} (m=1/2) = a(a+1)\,,~~ E_{a} (m=1/2) =
a(a+1)/2\,, \nonumber \\
&&\psi_{j}(x,m=1/2) \propto \psi_{2a-j} (y,m=1/2)\,.
\eea
On using 
\bea\label{3.6}
&&\dn(x,m)=\sqrt{1-m}\, \sn[ix+K'(m)+iK(m),1-m]\,, \nonumber \\
&&\sqrt{m}\,\cn(x,m)=i\sqrt{1-m} \,\cn[ix+K'(m)+iK(m),1-m]\,,
\eea
one can immediately and explicitly verify the relations (\ref{3.4})  and 
(\ref{3.5}) for $a=1,2,3,4$. 

For the $a$ = 2 case, 
the Lam\'{e} potential has 2 bound bands and a
continuum band.  
The energies and wave functions of the
five band edges are well known \cite{mw,ar}. The lowest energy band 
ranges from $2+2m-2\delta$ to $1+m$, the second energy band ranges 
from $1+4m$ to $4+m$ and the continuum starts at energy $2+2m+2\delta$, 
where $\delta = \sqrt{1-m+m^2}$. For subsequent application of the supersymmetric quantum mechanics formalism, it is convenient to have a potential $V_-(x)$ whose ground state energy is zero. This is easily accomplished via a re-defined potential with the ground state energy subtracted out. For the problem under consideration, one has $V_-(x)=6m\sn^2(x,m)-2-2m+2\delta$. This potential $V_-(x)$ is plotted in Fig. 1. The eigenstates of all the band edges 
are given in Table 1. Note that in the 
interval $2K(m)$ corresponding to the 
period of the  Lam\'{e} potential, the number of nodes 
increases with energy, in agreement with the oscillation theorem.
From Table 1, it is easy to check that, as expected, the relations
(\ref{3.4}) and (\ref{3.5}) are indeed true.

We might add here that in case $a$ is not an integer, then for Lam\'e 
potentials one has an infinite number of bands and band gaps for which (to the best of our
knowledge) no analytic results are available. However, when $a$ is
half-integral, 
then $(2a+1)/2$
mid-band levels are in principle analytically known and each of them is doubly
degenerate \cite{mw,ar}.

{\noindent\bf (b) Associated Lam\'{e} Potentials:}
We now expand our discussion to the band edges and wave functions of a 
much richer class of periodic potentials given by eq. (\ref{2})
called associated Lam\'e (AL) potentials, 
since the corresponding Schr\"odinger equation is called the associated 
Lam\'{e} equation \cite{mw}. More precisely, we often refer to the 
AL potential of eq. (\ref{2}) as the $(p,q)$ potential 
where $p=a(a+1)$ and $q=b(b+1)$ and note 
that $(p,0)$ potentials are just the ordinary Lam\'{e} potentials. 
The AL potentials (\ref{2}) can also be re-written in 
the alternative form $V(x)=pm~{\rm sn}^2(x)+qm~{\rm sn}^2(x+K(m))~$ 
\cite{gr,ks1}.
Clearly, the potentials 
$(p,q)$ and $(q,p)$ have the same energy spectra with wave functions shifted by $K(m)$. 
Therefore, it is sufficient to restrict our attention to $p \ge q$,
i.e. $a \ge b$.

In general, for any values of $p$ and $q$, the AL potentials 
have a period $2K(m)$, but
for the special case $p=q,$ the period is $K(m).$  
From a physical viewpoint, if one thinks of a Lam\'{e} potential $(p,0)$ as 
due to a one-dimensional regular array of atoms with spacing $2K(m)$ 
and ``strength" $p$, then the AL potential $(p,q)$ 
results from two alternating types of atoms spaced by $K(m)$ 
with ``strengths" $p$ and $q$ respectively. If the two types of 
atoms are identical [which makes $p=q$], one expects a potential of 
period  $K(m)$.

Extrema (defined for this discussion as either local or global maxima and 
minima) of AL potentials are easily found by 
setting $dV(x)/dx = 0.$  Extrema occur when ${\rm sn}(x)=0$, or ${\rm cn}(x)=0$. Also, for fixed values of $q$ and $m$, there are additional extrema if 
$p$ lies in the critical 
range $$q(1-m) \le p \le q/(1-m)~~.$$  
The AL potentials 
for $q=2, m = 0.5$ and several values of $p$ are plotted in Fig. 2. 
In the critical range of $p$ 
values $1 \le p \le 4 ~,$ one expects additional extrema, and these are 
clearly seen.

{\noindent \bf (c) Parabolas of Solvability:}
The AL equation is   
\be\label{3.7}
-\frac{d^2\psi}{dx^2} 
+ [a(a+1)m~{\rm sn}^2 (x) + b(b+1)m~{{\rm cn}^2(x)\over {\rm dn}^2 (x)}-E]\psi =0  ~~.
\ee
On substituting
$\psi (x) = [{\rm dn}(x)]^{-b} \ y(x) \, ,$
it is easily shown that $y(x)$,
satisfies the Hermite elliptic equation \cite{mw}.
On further substituting ${\rm sn}(x) =  \sin t \, , \ \  y(x) \equiv z (t) \, ,$
one obtains Ince's equation
\be\label{3.8}
(1+A\cos2t)z''(t)+B\sin2t Z'(t)+(C+D\cos2t)Z(t)=0\,,
\ee
where
\bea\label{3.9}
&&A=\frac{m}{2-m}\,,~~B=\frac{(2b-1)m}{2-m}\,,~~C=\frac{\lambda-(a+b)(a+1-b)m}
{2-m}\,, \nonumber \\
&&D=\frac{(a+b)(a+1-b)m}{2-m}\,,~~\lambda = E-mb^2\,,
\eea
which is a well known quasi exactly solvable equation \cite{mw}.
In particular if $a+b+1 = n$
( $n = 1,2,3,...$) then one obtains $n$ solutions, which are given in Table 2. 
In particular, for any given choice of $p = a(a+1),$ Table 2 lists the 
eigenstates of the 
AL equation for various values of $q=b(b+1)$. 

For $q=a(a-1)$, there is just one 
eigenstate with energy $ma^2$ and wave function $\psi = dn^a (x)$. Since the wave function 
has period $2K(m)$ and is nodeless, this is clearly the ground state wave function of the $(a(a+1),a(a-1))$ potential for any real choice of the parameter $a$. 
The equations $p=a(a+1)$ and $q=a(a-1)$ are the parametric forms of the equation of the 
parabola $(p-q)^2 = 2(p+q)$, which is plotted in Fig. 3 and labeled $P1$. For any point on the parabola, one knows the ground state wave function and energy $E_0 = ma^2$. The parabola $P1$ includes the points (2,0) and (6,2).

For $q=(a-1)(a-2)$, we see from Table 2 that two eigenstates with 
energies $1+m(a-1)^2$ and $1+ma^2$ are known. Since they have period 
$4K(m)$ and just one node in the interval $L=2K(m)$, they must correspond to 
the first and second band edge energies $E_1$ and $E_2$ of the 
$(a(a+1),(a-1)(a-2))$ potential. Eliminating $a$ from the equations 
$p=a(a+1)$ and $q=(a-1)(a-2)$ gives the  
``parabola of solvability" $(p-q)^2 = 8(p+q)-12$, which is plotted in Fig. 3 
and labeled $P2$.  This parabola includes the points (2,0) and (6,0) which 
correspond to Lam\'{e} potentials. Similarly, the parabolas of solvability 
$Pn~(n=0,1,2,...)$ corresponding to $q=(a-n+1)(a-n)$  in Table 2 are plotted. 
Note that $n$ 
eigenstates are known for any point on the parabola of solvability $Pn$.

All Lam\'{e} as well as AL potentials 
for which $a,b$ are unequal
integers, have two parabolas of solvability passing through. This
provides a good understanding of why these are completely solvable
problems. 
For instance, the $(2,0)$ potential is at the intersection of parabolas 
P1 (1 known state) and P2 (2 known states), thus giving the 3 known band edges.
Similarly, the $(6,2)$ AL potential. 
lies on parabola P1 (1 known nodeless state of period $2K(m)$) and parabola 
P4 (4 known states of period $4K(m)$, two with 1 node and two with 3 nodes). 
Since we know from the oscillation theorem that 2 states of period $2K(m)$ are 
missing, so it would appear that this is an example of a quasi exactly 
solvable 
potential. However, using a well known theorem about Ince's
equation \cite{mw} we discuss below that the two ``missing" states are degenerate and have no band gap between them. Hence, AL potentials
are also exactly solvable periodic problems with a finite number of
band gaps when both $a$ and $b$ are unequal positive integers. 

Other fully solvable examples correspond to AL potentials with $a=b=$ integer. 
For example, the $(2,2)$ 
potential has period $K(m)$. It lies on parabola P3 (3 known states) and the 
band edge periods are $K(m), 2K(m), 2K(m)$. We
shall discuss these examples in some detail in Sec. V where we show
that in view of the Landen transformation formulas for Jacobi elliptic functions, 
these problems are essentially related to Lam\'e potentials with integer 
$a$ and hence exactly solvable.\sss
  
{\noindent \bf (d) Exact Results:}
Several exact results are known in the literature about the Ince's 
equation (\ref{3.8}). In particular, it is well known that the system
satisfying Ince's eq. (\ref{3.8}) has {\it at most} $j+1$ band gaps of
period $\pi [2\pi$] in case the polynomial $Q(\mu) [Q^{*}(\mu)]$
defined by
\be\label{3.8a}
Q(\mu)=2A\mu^2 -B\mu-D/2\,,~~Q^{*}(\mu)=2A(\mu -1/2)^2 -B(\mu
-1/2)-D/2\,,
\ee
where $A,B,D$ are as given by eq. (\ref{3.9}) 
has either non-negative integral roots, the highest of
which is $j$, or negative integral roots, the lowest of which is  
$-j-1$. The fact that the Lam\'e potentials (\ref{1}) have only $a$
band gaps when $a$ is any positive integer is easily understood
from here. We now show \cite{ks2} that on applying this theorem, 
one can draw the following
conclusions about the associated Lam\'e potentials:

\begin{enumerate}

\item {\bf $a,b$ unequal positive integers}: In this case one can show
that for $a > b$, there are only $a$ bound bands followed by a
continuum band out of which the lowest $a-b$ bands are normal bands
with one band edge wave function of period $2K(m)$ and the other with 
period $4K(m)$, while the top $b$ bound bands are unusual in that both of their band
edges  are of period $2K(m) [4K(m)]$ in case $a-b$ is an even [odd]
integer. Further, all the $2a+1$ band edge eigenvalues and
eigenfunctions are in principle known analytically. As an
illustration, in Table 3 we have given all five band edge energy
eigenstates of the $(6,2)$ potential.

\item {\bf Both $a,b$ being half-integral}: In this case there are
infinite number of bands and band gaps. However, but for the lowest 
$a-b$ bands, the rest are unusual in that both of their band edges
are of period $2K(m) [4K(m)]$ in case $a-b$ is an even [odd] integer.
Further, in this case $a-b$ band edge energy eigenstates of period $2K(m)
[4K(m)]$ are analytically known in principle, in case $a-b$ is an odd [even]
integer. Besides, one also analytically knows the energy eigenvalues
and eigenfunctions of $b+1/2$ mid-band states of period $2K(m)
[4K(m)]$ each of which is doubly
degenerate in case $a-b$ is an odd [even] integer \cite{ks2}. 

\item {\bf Either $a+b$ or $a-b$ integral}:
When neither $a$ nor $b$ is integral or half-integral but either
their sum or difference is an integer, then one again has infinite
number of bands and band gaps. Further, one can show that if
either $a+b=2N$ or $a-b=2N+1$, then there are at most $N+1$ band gaps of
period $2K(m)$ and one analytically knows the energy eigenvalues
and eigenfunctions of $2N+1$ band edges of period $2K(m)$. On the other hand, 
if $a+b=2N+1$ or $a-b=2N$, then there are at most $N$
band gaps of period $4K(m)$ and one has analytical expressions
for $2N$ band edges of period $4K(m)$. 

\end{enumerate}

Finally, just as $a+1/2$ mid-band states, with each being doubly
degenerate, are analytically known in case parameter $a$ of the Lam\'e
potential (\ref{1}) is half integral, one can show that a similar number
of doubly degenerate mid-band states are also known for every integral
value of $b$ in the case of AL
potentials (\ref{2}) with half-integral $a$ \cite{ks2}.

\vspace{.2in}
{\noindent\bf IV. SUPERSYMMETRIC PARTNER POTENTIALS:}

\noindent{\bf (a) Lam\'{e} potentials:} 
Let us first apply the 
supersymmetric quantum mechanics formalism  as explained in Sec.
II to the Lam\'{e} 
potentials (\ref{1}) when $a$ is positive integer.
Since analytic solutions are known for integer values of $a$ \cite{ar}, 
the supersymmetric partner potentials can be readily computed using
eqs. (\ref{2.2}) and (\ref{2.2a}). 
We first discuss the results for small integer values of $a$, 
and then present some eigenstate results for arbitrary integer values 
of $a$. 

In order to use the supersymmetry formalism,
we must shift the Lam\'{e} potential by a constant to ensure
that the ground state i.e. (the lower edge of the lowest band) has energy
$E = 0$. For $a=1$, one has $V_{-} (x) = 2m {\rm sn}^2(x)-m, ~\psi_0^{(-)}={\rm dn} (x)$ and the superpotential is $W=m{\rm sn}(x) {\rm cn}(x)/{\rm dn}(x)$. The partner $V_+(x)$ turns out to be just $V_-(x-K(m))$, so that this is an example of self-isospectrality. For $a=2$, the potential is
\be\label{4.1}
V_{-} (x) = 6m {\rm sn}^2(x)-2-2m+2\delta~,~~\delta \equiv \sqrt{1-m+m^2}~,
\ee
with a unnormalized ground state wave function 
$\psi^{(-)}_{0} (x) = 1+m+\delta-3m {\rm sn}^2 (x)$ \cite{ar}.
The corresponding superpotential is
\be\label{4.2}
W = -{6m {\rm sn}(x) {\rm cn}(x) {\rm dn}(x)/
\psi^{(-)}_{0} (x)} \, ,
\ee
and hence the partner potential $V_{+} (x)$ for the potential $V_{-} (x)$ given in eq. 
(\ref{4.1}) is
\be\label{4.3}
V_{+}(x) = -V_-(x)+ 
{72 m^2 {\rm sn}^2(x) {\rm cn}^2(x) {\rm dn}^2(x)\over [1+m+\delta-3m {\rm sn}^2 (x)]^2}~~.
\ee
Although supersymmetry guarantees that the potentials $V_{\pm}$ 
are isospectral, in this example they are not
self-isospectral.
Therefore, $V_+(x)$ as given by eq. (\ref{4.3}) is a new 
periodic potential which is strictly isospectral
to the potential (\ref{4.1}) and hence it also has 2 bound bands and a 
continuum band. In Fig. 1 we have plotted the
potentials $V_{\pm}(x)$ corresponding to $a = 2$ for
$m = 0.8$. Using 
eqs. (\ref{2.7}) and (\ref{2.8})
and the known eigenstates of $V_{-} (x)$,
we can immediately compute all the band-edge Bloch wave functions for 
$V_{+} (x)$. In Table 1 
we have given the energy eigenvalues and wave functions for the isospectral 
partner potentials $V_{\pm} (x)$. 
In summary, for integral $a$, 
Lam\'e potentials with $a \ge 2$ are not self isospectral. They have distinct
supersymmetric partner potentials even though both potentials  
have the same $(2a+1)$ band edge eigenvalues.

\noindent{\bf (b)  Associated Lam\'{e} potentials:}  
It is easily checked from Table 2
that the solution corresponding to $q = a(a-1)$ as well as one of the 
$q = (a-2)(a-3)$ 
solutions are nodeless and correspond to the ground state.
Hence, for these cases, one can obtain
the superpotential and hence the partner potential $V_{+}$. 
For example, consider
the case of $p = a(a+1), q = a(a-1)$. In this case $W$ is given by
$
W= am {{\rm sn} (x) {\rm cn} (x)/ {\rm dn} (x)} \, ,
$
so that the corresponding partner potentials are
\be\label{4.4}
V_{\pm} = (a\pm 1)a m {{\rm cn}^2 (x)\over {\rm dn}^2 (x)}
+ m a (a \mp 1) {\rm sn}^2 (x)
- m a^2 ~~.\nonumber \\
\ee
These partner potentials are
self-isospectral and so supersymmetry yields   
nothing new. As an illustration,
 the potential (6,2) as discussed in Table 3 is a
self-isospectral potential and supersymmetry yields nothing new. 

Let us now consider the partner potential computed from the ground state for 
the $p=a(a+1), q= (a-2)(a-3)$
case.  Here
$\psi_0 (x) = [m(a-1)-1-\delta'+m(2a-1) {\rm sn}^2 (x)] ({\rm dn} (x))^{a-2}$,
where
$\delta' = \sqrt{1-m+m^2(a-1)^2}$.
The corresponding superpotential
$W$ turns out to be
\be\label{4.5}
W = {m(a-2) {\rm sn} (x) {\rm cn} (x)\over {\rm dn} (x)}
- {2m(2a-1) {\rm sn} (x) {\rm cn} (x) {\rm dn} (x)
\over [m(1-a)-1-\delta' +m(2a-1){\rm sn}^2 (x)]} \, .
\ee
Hence the corresponding partner potentials are
\be\label{4.6}
V_{-}  = ma(a+1) {\rm sn}^2 (x)
             +  m(a-3)(a-2){{\rm cn}^2 (x)\over {\rm dn}^2 (x)}
-2-m(a^2-2a+2)+2\delta',
V_{+} = - V_{-} +2 W^2.
\ee
These potentials are not self-isospectral.
Thus one has discovered
a whole new class of periodic potentials $V_{+} (x)$
for which  
three states are analytically known no matter what $a$ is. 
In particular, the energy
eigenfunctions of these three states are easily obtained by taking
the corresponding energy eigenstates of $V_{-}$ as given in Table 2 and 
using eqs. (\ref{2.7}) and (\ref{2.8}). 

We might add here that applying 
second order Darboux transformations 
to the Lam\'e potentials, 
Fern\'andez et al \cite{fer} have obtained an interesting nonlocal
effect, by which the transformed potential becomes an exact or approximately
displaced copy of the original one.

\vspace{.9in}
{\noindent\bf V. SUPERPOSITION OF LAM\'E AND ASSOCIATED LAM\'E
POTENTIALS}

Using the recently discovered cyclic identities \cite{ks5} and Landen
transformation formulas \cite{lau,ks4} for Jacobi elliptic functions, we shall
now further expand the list of analytically solvable periodic
potential problems. In particular, we obtain the exact band edge
eigenstates for potentials obtained via a certain specific type of linear superposition 
of Lam\'e as well as AL potentials. 

{\noindent \bf Superposition of Lam\'e potentials:} Consider
the following linear superposition of $p$ Lam\'e potentials with translated arguments:
\be\label{5.1}
V(x,m)=a(a+1)m\sum_{j=1}^{p} \sn^2 (x_{j},m)\,,  
\ee
where $x_j \equiv x+2(j-1)K(m)/p$.
Note that for the special case of
$p=2$, this corresponds to the AL potential with $a=b$. On expressing
$\sn^2(x,m)$ in terms of $\dn^2(x,m)$, the
corresponding Schr\"odinger equation is given by
\be\label{5.2}
-\psi''(x)-
\big [a(a+1)\sum_{j=1}^{p} \dn^2 (x_{j},m) \big
] \psi (x)= [E-a(a+1)p] \psi (x)\,.  
\ee 
On making use of the Landen transformation formula \cite{lau,ks4}
\be\label{5.3}
\sum_{j=1}^{p} \dn
(x_{j},m)=\frac{1}{\alpha}\dn(\frac{x}{\alpha},\tilde{m})\,,
\ee
where
\be\label{5.4} 
\frac{1}{\alpha} = 
\sum_{j=1}^{p} \dn (2(j-1)K(m)/p,m) \,,
\ee
and
\be\label{5.5}
\tilde{m}=(m-2)\alpha^2
+2\alpha^3 \sum_{j=1}^{p} \dn^3 (2(j-1)K(m)/p,m) \,,
\ee
one can rewrite the Schr\"odinger eq. (\ref{5.2}) in the form
\be\label{5.6}
-\psi''(x)+
\frac{a(a+1)\tilde{m}}{\alpha^2} \sn^2 (\frac{x}{\alpha},\tilde{m}) 
 \psi (x) = \left [E-a(a+1)(p+2A_{d}-\frac{1}{\alpha^{2}}) \right ] \psi (x) \,.  
\ee 
Here $A_d \equiv \sum_{j < k =1}^{p} \dn(x_j,m) \dn(x_k,m)$. 
Using recently discovered cyclic identities, $A_d$ can be shown to be
\cite{ks5}
\bea\label{5.7}
&&A_d =\frac{p}{2}\sqrt{1-m}
+p \sum_{j=1}^{(p-2)/2} \big [\dn (2jK(m)/p,m)-\cs (2jK(m)/p,m)
Z(2jK(m)/p,m) \big ]\,,~p~{\rm even},\nonumber \\
&&=p \sum_{j=1}^{(p-1)/2} \big [\dn (2jK(m)/p,m)-\cs (2jK(m)/p,m)
Z(2jK(m)/p,m) \big ]\,,~p~{\rm odd},
\eea
where $\cs(x,m) \equiv \frac{\cn(x,m)}{\sn(x,m)}$, while $Z(x,m)$ is
the Jacobi zeta function \cite{gr}. 
On making the transformation $x=\alpha y$, the Schr\"odinger eq.
(\ref{5.6}) can be rewritten as a Lam\'e equation but for the modulus parameter
$\tilde{m}$ and with energy $E^{(L)} (\tilde{m})$:
\be\label{5.8}
-\psi''(y)+
\big [a(a+1)\tilde{m} \sn^2 (y,\tilde{m}) \big
] \psi (y) = E^{(L)} (\tilde{m}) \psi (y) \,.  
\ee 
Here the true energy eigenvalues $E(m)$ of the superposed potentials
(\ref{5.1}) are related to the eigenvalues $E^{(L)} (\tilde{m})$ 
of the Lam\'e potentials by
\be\label{5.9}
E_j (m) = \frac{E_j ^{(L)} (\tilde{m})}{\alpha^2}
+a(a+1)[p+2A_d-\frac{1}{\alpha^2}]\,,
\ee
where $\alpha$ and $A_d$ are as given by eqs. (\ref{5.4}) and
(\ref{5.7}) respectively while $\tilde{m}$ and $m$ are related by 
eq. (\ref{5.5}). Similarly, the true eigenfunctions of the superposed 
potentials (\ref{5.1}) are related to those of eq. (\ref{5.8}) by
\be\label{5.10}
\psi_j (x,m) \propto \psi_j ^{(L)} (\frac{x}{\alpha},\tilde{m})\,.
\ee

As an illustration, for $p=2$ (i.e. $a=b$ AL case), 
it is well known \cite{lau,ks4} that
\be\label{5.11}
\alpha=\frac{1}{1+\sqrt{1-m}}\,,~~A_d = \sqrt{1-m}\,,~~\tilde{m}=
\frac{(1-\sqrt{1-m})^2}{(1+\sqrt{1-m})^2}\,,
\ee
and hence the energy eigenvalues of the $a=b$ AL potentials
(\ref{5.1}) (with $p=2$) are given in terms of those of Lam\'e
potential (\ref{1}) (but with the modulus parameter $\tilde{m}$) by 
\be\label{5.12}
\alpha^2 E^{(AL)}_j (m)= E_j ^{(L)} (\tilde{m})+a(a+1)
\sqrt{\tilde{m}}\,.
\ee
For $a=1,2$ we can immediately verify that the energy eigenvalues and
eigenfunctions of the  
(2,2) and (6,6) potentials as calculated from here are identical (as
they should be) to those calculated by us by an entirely different
method where no use of either Landen transformations or the cyclic
identities was made \cite{ks1}.

In fact we can relate the energy eigenstates of the $a=b$ AL
potentials at different modulus parameters by using the relationship
(\ref{3.4}) between the corresponding Lam\'e energy eigenstates. 
For this purpose, it is best to reexpress the relation (\ref{5.12}) 
entirely in terms of $\tilde{m}$ by noting that on using eq.
(\ref{5.11}) we have
\be\label{5.12a}
m=\frac{4\tilde{m}}{(1+\sqrt{\tilde{m}})^2}\,.
\ee
Using eqs. (\ref{5.11}) to (\ref{5.12a}) in eq. (\ref{3.4}) we find
that the AL eigenvalues are related by (where for simplicity we have
replaced $\tilde{m}$ everywhere by $m$)
\be\label{5.12b} 
\frac{(1+\sqrt{m})^2}{4} E_j^{AL}(m_1)+
\frac{(1+\sqrt{1-m})^2}{4} E_{2a-j}^{AL}(m_2)
=a(a+1)[1+\sqrt{m}+\sqrt{1-m}]\,,~~j=0,1,...,2a\,,
\ee
where
\be\label{5.12c}
m_1 = \frac{4\sqrt{m}}{(1+\sqrt{m})^2}\,,~~
m_2 = \frac{4\sqrt{1-m}}{(1+\sqrt{1-m})^2}\,.
\ee
For $a=1,2$ we immediately verify that these relations are indeed
true.
Further, the corresponding eigenfunctions are related by
\be
\psi_j (x,m_1) \propto \psi_{2a-j} (y,1-m)\,,
\ee
where $y = \frac{2ix}{1+\sqrt{m}}+K'(m)+iK(m)$.
 
Similarly, for $p=3$, it is well known that  
\be\label{5.13}
\alpha=\frac{1}{1+2q}\,,~~A_d = q(q+2)\,,~~\tilde{m}=m
\frac{(1-q)^2}{(1+q)^2(1+2q)^2}\,,~~q \equiv \dn(2K(m)/3,m)\,,
\ee
and hence the energy eigenvalues of (\ref{5.1}) (with $p=3$) are given
in terms of those of Lam\'e's with parameter $\tilde{m}$ by
\be\label{5.14}
E(m)=(1+2q)^2 \tilde{E} (\tilde{m})+2a(a+1)(1-q^2)\,.
\ee

Thus by using Landen transformations and recently
discovered cyclic identities one has discovered new solvable periodic
potentials. On following the last section, it is easy to see
that for any integer $a \ge 2$, corresponding to every superposed
potential, supersymmetry will give us another
new exactly solvable periodic potential. 

{\noindent \bf Superposition of Associated Lam\'e potentials:} Consider the
the following superposition of AL potentials
\be\label{5.15}
V(x,m)=a(a+1)m\sum_{j=1}^{p} \sn^2 (x_{j},m)  
+b(b+1)m\sum_{j=1}^{p} \sn^2 (x_{j}+K(m),m)\,,  
\ee
where $x_j \equiv x+2(j-1)K(m)/p$. Now proceeding exactly as above and
using the Landen transformation (\ref{5.3}) and the cyclic identity
(\ref{5.7}), one can show that the energy eigenvalues and
eigenfunctions of the superposed AL potentials (\ref{5.15}) are
related to those of AL potential (\ref{2}) (but with modulus parameter
$\tilde{m}$) by
\be\label{5.16}
E(m) = \frac{\tilde{E} (\tilde{m})}{\alpha^2}
+[a(a+1)+b(b+1)][p+2A_d-\frac{1}{\alpha^2}]\,,
\ee
\be\label{5.17}
\psi(x,m) \propto \psi(\frac{x}{\alpha},\tilde{m})\,,
\ee
where $\alpha,\tilde{m}$ and $A_d$ are given by eqs. (\ref{5.4}),
(\ref{5.5}) and
(\ref{5.8}) respectively.
Thus we have further enlarged the list of new exactly solvable periodic
problems. Needless to say that using supersymmetry, one can further
enlarge this list by obtaining partner potentials of the superposed
potentials which will have the same band
spectra.

As an illustration, for $p=2,3$, the energy eigenvalues of the
superposed potentials (\ref{5.15}) are given below in terms of the
corresponding AL potentials (with modulus parameter $m$ changed to 
$\tilde{m}$): 
\be\label{5.18}
E(m)=[1+\sqrt{1-m}]^2 \tilde{E} (\tilde{m})+m[a(a+1)+b(b+1)]\,,
\ee
\be\label{5.19}
E(m)=(1+2q)^2 \tilde{E} (\tilde{m})+2[a(a+1)+b(b+1)](1-q^2)\,.
\ee
where use has been made of eqs. (\ref{5.11}) and (\ref{5.13}).

\vspace{.2in}
{\noindent\bf VI. PT INVARIANT POTENTIALS WITH REAL BAND SPECTRA} 

We shall now show that the complex 
PT-invariant periodic potentials obtained by applying the anti-isospectral
transformation \cite{kra} $x 
\rightarrow ix+\beta$ (where $\beta$ is a nonzero real number)
to the
Lam\'e and AL potentials with a finite number of band gaps are also 
exactly solvable and have a finite number of band gaps.  
We give explicit expressions for the band edges of some of these 
potentials \cite{ks3}.
For the special case of the potential
$V(x)=-2m\sn^2(ix+\beta,m)$, we analytically obtain the
corresponding dispersion relation.

Let us begin with the simple observation that if $\psi (x)$ is a 
solution of the Schr\"odinger equation for the real potential 
$V(x)$ with energy $E$, then $\psi (ix+\beta)$ is a solution of the
Schr\"odinger equation for the complex potential $-V(ix+\beta)$ with
energy $-E$. The new potential $-V(ix+\beta)$, generated by the
anti-isospectral transformation $x \rightarrow ix+\beta$, is clearly
PT-symmetric and will be denoted by $V^{(PT)} (x)$. 

It is important to understand that the key point in obtaining these
results is that unlike
trigonometric and most other periodic functions, the Jacobi elliptic
functions are doubly periodic functions. This allows both 
$V(x)$ and $V^{(PT)} (x)$ to be simultenously periodic, even though the
periods are different. Note that the arbitrary nonzero constant 
$\beta$ in the transformation $x \rightarrow
ix+\beta$ helps us to avoid the singularities of Jacobi elliptic
functions. 

Let us first apply our approach to the Lam\'e potentials (\ref{1}). 
On applying the anti-isospectral transformation $x \rightarrow ix+\beta$ 
\cite{kra}, where $\beta$ is any non-zero real number, it is easily
shown that the band-edge eigenvalues and eigenfunctions of the
PT-invariant potentials 
\be\label{6.1}
V^{(PT)}(x)=-a(a+1)m\sn^2(ix+\beta,m)\,,~~a=1,2,3,...\,, 
\ee
are related to those of the Lam\'e potentials (\ref{1}) by \cite{ks3}
\be\label{6.2}
E_j^{(PT)}(m) =
-E_{2a-j}(m)\,,~~\psi_j^{(PT)}(x,m) 
\propto \psi_{2a-j}(ix+\beta,m)~,~~j=0,1,...
,2a\,.
\ee
It may be noted here that the PT-invariant complex potential (\ref{6.1})
is a periodic potential with period $2K'(m) (\equiv 2K(1-m))$. 
Thus we have shown   
that the PT-invariant periodic potential (\ref{6.1}) too has
precisely $a$ band gaps and hence $2a+1$ band edges at energies 
given by relation (\ref{6.2}). Special mention may be made of the
remarkable fact that for any integer $a$, all  bands and band gaps 
exchange their role as one goes from a Lam\'e potential 
to its PT-transformed version.

For any band structure problem, one important quantity is the
discriminant $\Delta$ \cite{mw} which gives information
about the number of band gaps as well as their widths. 
The question is if we can relate $\Delta^{(PT)}$ with
the discriminant $\Delta$ 
for the corresponding Lam\'e potential. We now show that using eqs.
(\ref{3.4}) and (\ref{6.2}) this is indeed possible. In particular, on
using these two equations we deduce that
\be\label{6.8}
E^{(PT)}_j (m) = E_{j} (1-m) -a(a+1)\,,~~j=0,1,...,2a\,,
\ee
and hence the corresponding discriminants are  
related by \cite{ks3} 
\be\label{6.9}
\Delta^{(PT)} (E,m)=\Delta[E+a(a+1),1-m]\,. 
\ee

As an illustration, in Fig. 4 we plot the real and
imaginary parts of the PT-invariant complex potential 
$-6m\sn^2 (ix+\beta,m)$.  
Using the well known results for the Lam\'e 
potential with $a=2$ (see Table 2) and eq. (\ref{6.2}), 
the ground state (lowest band edge)
eigenvalue and eigenfunction is easily shown to be
\be\label{6.10}
\psi_g(x)=\sn(ix+\beta,m)[1+m-\delta-3m\sn^2(ix+\beta,m)]\,,~~  
E_g = -2-2m-2\delta\,,
\ee
where $\delta=\sqrt{1-m+m^2}$.
In Table 4 we have given all the five  
band edge eigenvalues and eigenfunctions of this PT-invariant
potential, 
where we have subtracted off
the ground state energy from the potential so that the lowest band
edge by construction is at zero energy. Observe from the table that
the band edges are both periodic as well as anti-periodic with periods
of $2K'(m)$ and $4K'(m)$ respectively thereby showing that contrary to
the suggestion of Bender et al. \cite{bdm}, the absence
of anti-periodic band edges is not a general property of PT-invariant 
periodic potentials. 

Additional analytically solvable finite band gap potentials can be
obtained from 
here by using supersymmetry. In particular, consider the potential $V_-^{(PT)}(x)=-6m\sn^2(ix+\beta,m)-E_g$. Its ground state
eigenfunction 
is given in eq. (\ref{6.10}), and 
we find that the corresponding superpotential is
\be\label{6.11}
W^{(PT)}(x) = -i\frac{\cn(ix+\beta,m) \dn(ix+\beta,m)}{\sn(ix+\beta,m)}
+ 6im \frac{\cn(ix+\beta,m) \sn(ix+\beta,m) \dn(ix+\beta,m)}
{[1+m-\delta-3m\sn^2(ix+\beta,m)]}\,.
\ee
Thus the supersymmetric partner potential 
is   
\be\label{6.12}
V^{(PT)}_{+} (x) = [W^{(PT)}]^2 (x) +[W^{(PT)}]' (x)\,,
\ee
and it has the same energy spectrum as $V_-^{(PT)}(x)$
[see Table 4]. 
In this way, one has discovered yet another PT-invariant complex
potential with a finite number of band gaps. The corresponding band edge energy
eigenfunctions are immediately obtained from those of
$V^{(PT)}_{-}$ by using supersymmetry transformations given in Sec.
II.  

Yet another complex PT-invariant potential having the same band edge
eigenvalues as above is obtained by starting from the partner potential
$V_{+}(x)$ of the $a=2$ Lam\'e potential [eqs. (\ref{4.3})
and (\ref{4.1})] and then obtaining the corresponding PT-invariant
potential by the anti-isospectral transformation $x \rightarrow ix+\beta$.
Thus one has found three distinct complex PT-invariant periodic
potentials having the same band spectra. One can show that this is
true for any integral $a \ge 2$. On the other hand, for $a=1$ one can
show \cite{ks3} that the three potentials are not distinct but are 
self-isospectral (i.e. differ from each other by a translation by a
constant) \cite{df,fer}.
  
For the special case of the $a=1$ Lam\'e potential, the dispersion relation is
analytically known \cite{ww}. We
now show  \cite{ks3} that it is possible to derive the dispersion relation for the  
PT-invariant potential (\ref{6.1}) with $a=1$. To that
end we start from the Schr\"odinger equation: 
\be\label{6.13}
-\psi''(x)+[1+m-2m\sn^2(ix+\beta,m)] \psi(x) = E \psi(x)\,,
\ee
where we have subtracted the ground state energy from the potential so
that the new potential has zero ground state energy. On substituting
$y=ix+\beta$, eq. (\ref{6.13}) takes the form
\be\label{6.14}
-\psi''(y)+[2m\sn^2(y,m)-m] \psi(y) = (1-E) \psi(y)\,.
\ee
Now it is well known that two independent solutions of this equation
are given by \cite{ww}
\be\label{6.15}
\psi(x)=\frac{H(ix+\beta\pm \alpha_1)\exp[\mp (ix+\beta) Z(\alpha_1)]}
{\theta(ix+\beta)}\,,
\ee
where $H,\theta,Z$ are Jacobi eta, theta and zeta functions while
$\alpha_1$ is related to $E$ of eq. (\ref{6.13}) by 
\be\label{6.16}
E=m\sn^2(\alpha_1,m)\,.
\ee
On using the fact that while 
$\theta(ix+\beta)$ is periodic function with period 
$2K'(m)$, $H(ix+\beta)$ is only quasi-periodic \cite{ww}, i.e.
\be\label{6.17}
H(i[x+2K'(m)]+\beta)=H(ix+\beta)\exp[-\pi K'(m) /K(m)]\,,
\ee
and using the Bloch condition, it is easily shown that for the
PT-invariant complex potential (\ref{1}) with $a=1$, the dispersion relation is
given by
\be\label{6.18}
k = \mp \frac{\pi}{2K'(m)} \pm iZ(\alpha_1) + i\frac{\pi}{2K(m)}\,,
\ee
where $\alpha_1$ is given by eq. (\ref{6.16}). 

We now turn to the associated Lam\'e potentials given by eq.
(\ref{2}) where without any loss of generality we consider $a > b$ with
both being positive integers. 
On using the anti-isospectral transformation, it is easy to
see that the band edges of the potential (\ref{2}) and
its PT-transformed one as given by  
\be\label{6.19}
V(x)=-a(a+1)m\sn^2(ix+\beta,m)-b(b+1)m\frac{\cn^2 (ix+\beta,m)}
{\dn^2(ix+\beta,m)}\,,
\ee
are again related by the relation (\ref{6.2}). 

For the special case $a=b$, using relations (\ref{5.12a}),(\ref{5.12b}) and (\ref{6.2}) we obtain an interesting relation betwen the eigenvalues of the AL potential and the corresponding complex PT-invariant potential given by 
\be\label{6.20}
E_j^{(PT)}(m_1) =(\frac{1+\sqrt{1-m}}{1+\sqrt{m}})^2 E_j (m_2) -4a(a+1)\frac{(1+\sqrt{m}+\sqrt{1-m})}{(1+\sqrt{m})^2}\,,
\ee
where $m_{1,2}$ are as defined by eq. (\ref{5.12c}).
Hence we can immediately relate the discriminant $\Delta$ for the $a=b$ AL potential and the corresponding complex PT-invariant potential, i.e. we have 
\be\label{6.21} \Delta^{(PT)} (E,m_1) = \Delta[(\frac{1+\sqrt{m}}{1+\sqrt{1-m}})^2
(E+\frac{4a(a+1)(1+\sqrt{m}+\sqrt{1-m})}{(1+\sqrt{m})^2}, m_2]\,. 
\ee

\vspace{.2in}
{\noindent \bf VII. UNUSUAL BAND STRUCTURE FOR THE DOUBLE SINE-GORDON EQUATION} 

Finally, we would like to consider the band spectra of the double
sine-Gordon equation characterized by the periodic potential
\be\label{7.1}
V(x)=b^2 \sin^2 2x +2ab \cos 2x\,,
\ee
where $a,b$ are real, and without loss of generality $a$ is taken to be 
nonnegative. This equation arises in several areas of
condensed matter physics. It is well known that this system has an 
infinite number of bands and band gaps. Further, it is also well known
that if $a$ is an integer then $a$ band edges of period $\pi$($2\pi$) 
are analytically known depending on whether $a$ is an odd(even) 
integer \cite{raz,km}. However, what is not so well
known \cite{mw} is that in case $a$ is an integer, then the potential
has unusual band spectra. 

We start from the Schr\"odinger equation for the potential
(\ref{7.1}). On using the ansatz
\be\label{7.2}
\psi (x) = \exp(-b\cos2x /2) \phi (x)\,,
\ee
in the Schr\"odinger equation, it is easily shown that $\phi$ satisfies
Ince's eq. (\ref{3.8}) 
\be\label{7.3}
\phi''(x)+2b\sin 2x~ \phi'(x)+[E-2b(a-1)\cos2x]\phi(x) = 0\,.
\ee
As discussed before, this
is a quasi-exactly solvable equation, i.e. $a$ band edges of
period $\pi (2\pi$) are analytically known when $a$ is an odd
(even) integer \cite{raz,km}.
We can now compute the polynomials $Q(\mu)$ and $Q^{*}(\mu)$ 
defined in eq. (\ref{3.8a}), and from them we conclude  
that if $a$ is an odd (even) integer then there are at most
$\frac{a+1}{2} (\frac{a+2}{2})$ band gaps of period $\pi (2\pi$).

It is worth pointing out here that similar conclusions can also be
drawn about the complex PT-invariant periodic potential 
\be\label{7.4}
V(x)=-b^2 \sin^2 2x +2iab \cos 2x\,.
\ee
Note that this potential is invariant under the parity 
transformation $x \rightarrow x+\pi /2$ followed by $T$. As has been
shown in ref. \cite{km}, when $a$ is an even integer then
PT symmetry is spontaneously broken and there are no quasi-exactly 
solvable states of
period $2\pi$ since in this case all the eigenvalues are now complex
conjugate pairs. However, if $a$ is an odd integer then there are $a$ 
quasi-exactly solvable states of period $\pi$ and in this case PT symmetry is not
spontaneously broken. On completing the above analysis, it
follows that if $a$ is an odd integer then there are at most
$(a+1)/2$ band gaps of period $\pi$. Thus in this case the band
spectrum is rather unusual in that the majority of the infinite number of
band edges are anti-periodic and have period $2\pi$, thereby again 
showing that the absence of anti-periodic band edges is not 
a characteristic feature of PT-invariant potentials.

Acknowledgements: It is a pleasure to thank the U.S. Department of Energy for partial support of this research.

\newpage

\oddsidemargin      -0.5in
\noindent {\bf Table 1:} The eigenvalues and eigenfunctions for the 5 band edges corresponding
to the $a = 2$ Lam\'{e} potential $V_{-}$ as given by eq. (\ref{4.1}) 
which gives $(p,q)=(6,0)$ 
and its SUSY partner $V_{+}$. Here $B\equiv
1+m+\delta$ and $\delta\equiv \sqrt{1-m+m^2}$. The potentials $V_\pm$ have 
period $2K(m)$ and their analytic forms are given by eqs. (\ref{4.1}) 
and (\ref{4.3})
respectively.  The periods of various eigenfunctions and the number of nodes 
in the interval $2K(m)$ are tabulated.\sss

\bigskip
\begin{tabular}{cccccc}
\hline
 $E$ & $\psi^{(-)}$ & $[B-3m~{\rm sn}^2 (x)]\psi^{(+)}$ & 
${\rm Period}$ & ${\rm Nodes}$\\
\hline
 $0$ & $m + 1 +\delta-3m {\rm sn}^2 (x)$  & $1$  & $2K$ & $0$\\
 $ 2\delta-1-m$ & ${\rm cn}(x) {\rm dn}(x)$
& ${\rm sn}(x)[6m-(m+1)B+m {\rm sn}^2 (x) (2B-3-3m)]$ & $4K$ & $1$\\
 $ 2\delta-1+2m$ & ${\rm sn}(x) {\rm dn}(x)$
& $ {\rm cn}(x) [B+m {\rm sn}^2 (x) (3-2B)]$   & $4K$ & $1$\\
 $ 2\delta+2-m$ & ${\rm sn}(x) {\rm cn}(x)$
& ${\rm dn}(x) [B+{\rm sn}^2 (x) (3m-2B)]$   & $2K$ & $2$\\
 $4\delta$ & $m + 1 -\delta-3m {\rm sn}^2 (x)$ & ${\rm sn}(x)
{\rm cn}(x){\rm dn}(x)$   & $2K$ & $2$\\
\hline
\end{tabular}
\bigskip

\sss\sss\sss\sss
\oddsidemargin      -0.1in
\noindent {\bf Table 2:} Eigenvalues and
eigenfunctions for various associated Lam\'{e} potentials $(p,q)$
with $p = a(a+1)$ and $q = (a-n+1)(a-n)$ for $n = 1,2,3,...$. The periods of various eigenfunctions and the number of nodes in the interval $2K(m)$ are tabulated. Here $~\delta_4 \equiv \sqrt{1-m+m^2(a-1)^2}~$,$~ 
\delta_5 \equiv
\sqrt{4-7m+2ma+m^2(a-2)^2}~$,$~\delta_6 \equiv \sqrt{4-m-2ma+m^2(a-1)^2}~$,
$~\delta_7 \equiv \sqrt{9-9m+m^2(a-2)^2}~.$ \sss

\bigskip
\begin{tabular}{ccccccc} 
\hline
 $q$ & $E$ & ${\rm dn}^{-a} (x) \psi $ & ${\rm Period}$ & ${\rm Nodes}$\\
\hline
 $a(a-1)$ & $ma^2$  & $1$ & $2K$ & $0$\\
 $(a-1)(a-2)$ & $1+m(a-1)^2$ & ${{\rm cn}(x)\over {\rm dn}(x)}$ & $4K$ & $1$\\ 
 $(a-1)(a-2)$ & $1+ma^2$ 
& $ {{\rm sn}(x)\over {\rm dn}(x)}$ & $4K$ & $1$\\
 $(a-2)(a-3)$ & $2+m(a^2-2a+2)\pm 2\delta_4$ 
& ${[m(2a-1) {\rm sn}^2 (x)-1+m-ma
\pm\delta_4]\over {\rm dn}^2 (x)}$ & $2K$ & $2,0$\\
 $(a-2)(a-3)$ & $4+m(a-1)^2$ & 
${{\rm sn}(x) {\rm cn}(x)\over {\rm dn}^2 (x)}$ & $2K$ & $2$\\
 $(a-3)(a-4)$ & $5+m(a^2-4a+5)\pm 2\delta_5$ 
& ${{\rm cn}(x)[m(2a-1) {\rm sn}^2 (x)-2+2m-ma \pm \delta_5]
\over {\rm dn}^3 (x)}$ & $4K$ & $3,1$\\
 $(a-3)(a-4)$ & $5+m(a^2-2a+2)\pm 2\delta_6$ 
& ${{\rm sn}(x)[m(2a-1) {\rm sn}^2 (x)-2+m-ma\pm \delta_6]
\over {\rm dn}^3 (x)}$ & $4K$ & $3,1$\\
$(a-4)(a-5)$ & $10+m(a^2-4a+5)\pm2\delta_7$ & ${{\rm sn} (x) {\rm cn} (x) 
[m(2a-1){\rm sn}^2 (x)-3+2m-ma\pm\delta_7 ] \over {\rm dn}^4 (x)}$ & $2K$ & $4,2$\\
\hline
\end{tabular}
\bigskip

\newpage

\vskip 2.2 true cm
\noindent {\bf Table 3:} The five eigenvalues and
eigenfunctions for the self-isospectral associated Lam\'{e} potential corresponding to $a=2, b=1$ which gives
$(p,q) = (6,2)$. The potential is 
$V_{-} (x)= 6m {\rm sn}^2 (x) +2m {{\rm cn}^2 (x)
\over {\rm dn}^2 (x)} -4m$, and has period $2K(m).$ The number of nodes in the interval $2K(m)$ is tabulated. \sss

\bigskip
\begin{tabular}{cccccc}
\hline
 $E$ & $\psi^{(-)}$ & 
${\rm Period}$ & ${\rm Nodes}$\\
\hline
 $0$ & $\dn^2 (x)$  
 & $2K$ & $0$\\
 $ 5-3m-2\sqrt{4-3m}$ & ${\cn (x)\over \dn (x)} [3m{\rm sn}^2 (x) -2
-\sqrt{4-3m}]$
 & $4K$ & $1$\\
 $ 5-2m-2\sqrt{4-5m+m^2}$ & ${{\rm sn}(x) \over {\rm dn}(x)}
[3m{\rm sn}^2 (x)-2-m-\sqrt{4-5m+m^2}]$
 & $4K$ & $1$\\
 $ 5-2m+2\sqrt{4-5m+m^2}$ & ${{\rm sn}(x) \over {\rm dn}(x)}
[3m{\rm sn}^2 (x)-2-m+\sqrt{4-5m+m^2}]$
 & $4K$ & $3$\\
 $ 5-3m+2\sqrt{4-3m}$ & ${\cn (x)\over \dn (x)} [3m{\rm sn}^2 (x) -2
+\sqrt{4-3m}]$
 & $4K$ & $3$\\
\hline
\end{tabular}
\bigskip

\vskip 2.2 true cm

\sss\sss\sss\sss
\oddsidemargin      -0.1in
\noindent {\bf Table 4:} The eigenvalues and eigenfunctions for the 5 band edges corresponding
to the PT-invariant potential $V_{-}^{(PT)} =-6m\sn^2(ix+\beta,m)-E_g$
where $E_g=-2-2m-2\sqrt{1-m+m^2}$ [eq. (\ref{6.10})].
The potential has period $2K'(m)$. 
The periods of various eigenfunctions are also tabulated. \sss

\bigskip
\begin{tabular}{cccccc}
\hline
 $E$ & $\psi^{(-)}$ &  
${\rm Period}$\\
\hline
 $0$ & $1+m -\delta -3m {\rm sn}^2 (ix+\beta,m)$
& $2K'(m)$\\
 $ m - 2+2\delta$ & ${\rm cn}(ix+\beta,m) \sn(ix+\beta,m)$
& $4K'(m)$ \\
 $ 1 - 2m+2\delta$ & ${\rm sn}(ix+\beta,m) \dn(ix+\beta,m)$
& $4K'(m)$ \\
 $ 1 + m+2\delta$ & ${\rm cn}(ix+\beta,m) \dn(ix+\beta,m)$
& $2K'(m)$ \\
 $4\delta$ & $1+m +\delta-3m {\rm sn}^2 (ix+\beta,m)$
& $2K'(m)$\\
\hline
\end{tabular}
\bigskip

\newpage
\begin{figure}[ht] \label{fig1}
    \centering
    \epsfig{file=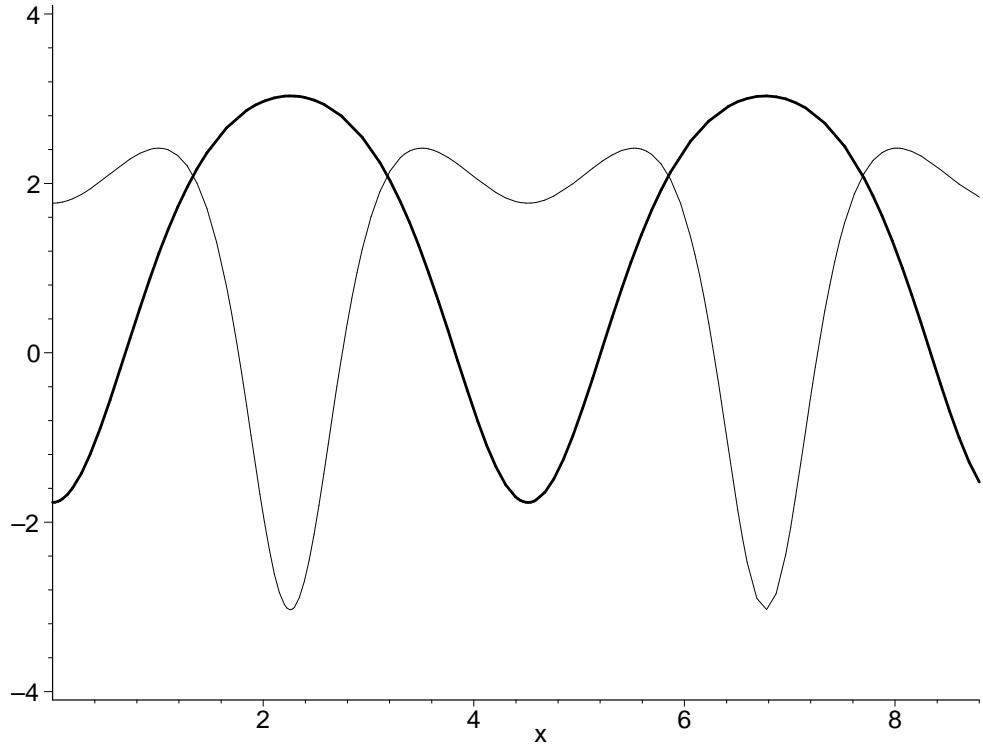,clip=,width=5.2in}
\caption{The (6,0) Lam\'{e} potential $V_-(x)=6m\sn^2 (x,m)-2-2m+2 \sqrt{1-m+m^2}$ [thick line] and its supersymmetric partner potential $V_{+} (x)$ [thin line] 
as given by eq. (\ref{4.3}). The curves are plotted for the choice $m=0.8~$. Both $V_-(x)$ and $V_{+} (x)$ have their lowest band edge at zero energy.}
\end{figure}
\begin{figure}[ht] \label{fig2}
    \centering
    \epsfig{file=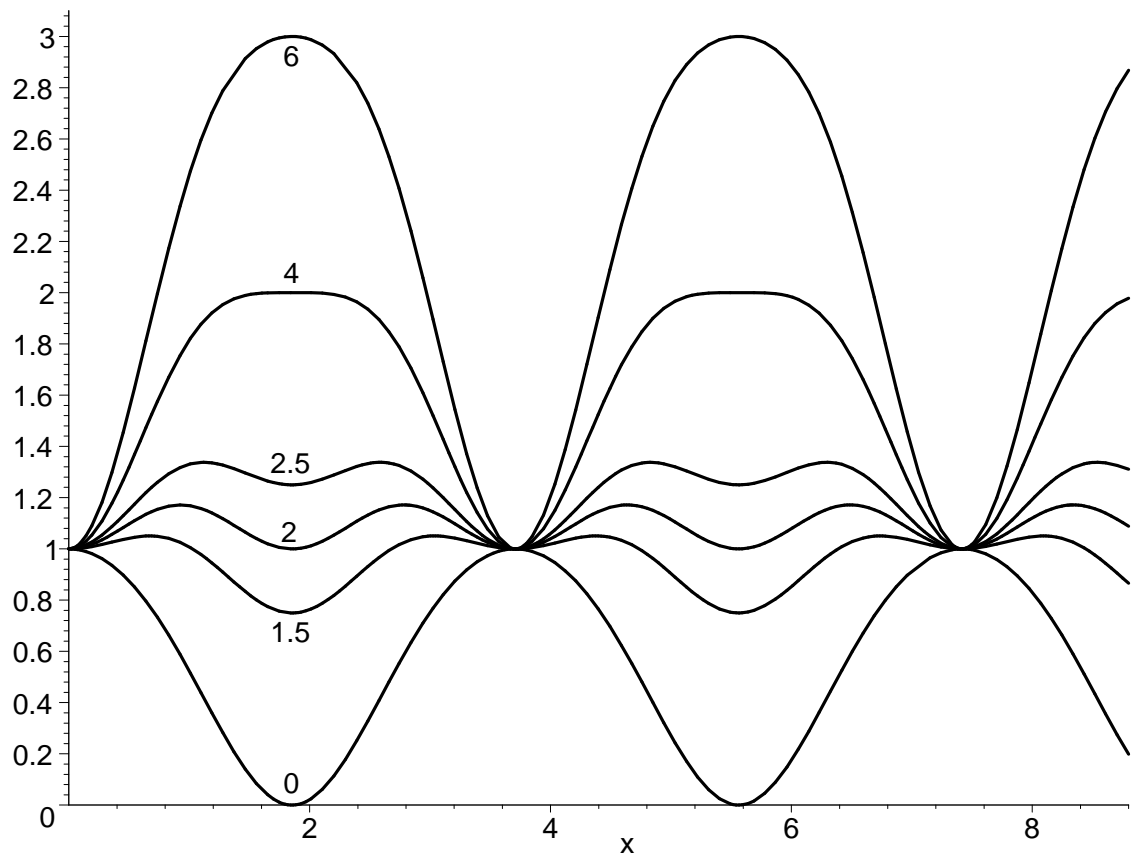,clip=,width=6.2in}
\caption{Plots of the $(p,q)$ associated Lam\'{e} potentials for $q=2, ~m = 0.5$ and several values of $p$.}
\end{figure}
\begin{figure}[ht] \label{fig3}
    \centering
    \epsfig{file=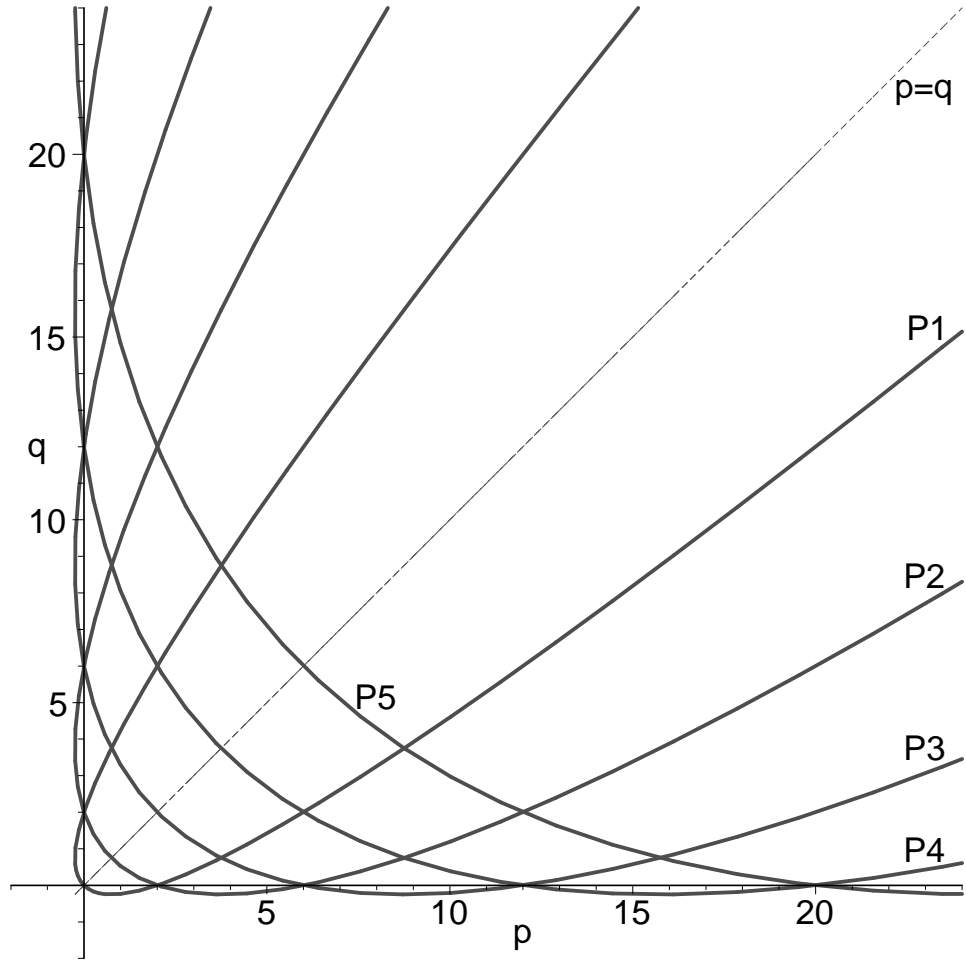,clip=,width=5.2in}
\caption{Parabolas of solvability.  This figure illustrates all associated 
Lam\'{e} potentials $(p,q)$ which are quasi solvable. Each parabola 
corresponds to a choice of $q$ in Table 2. Parabola $Pn$ is for 
$q=(a-n+1)(a-n)$ for $n=1,2,3,...$. One knows $n$ eigenstates for any 
point on parabola $Pn$ from Table 2. }
\end{figure}
\begin{figure}[ht] \label{fig4}
    \centering
    \epsfig{file=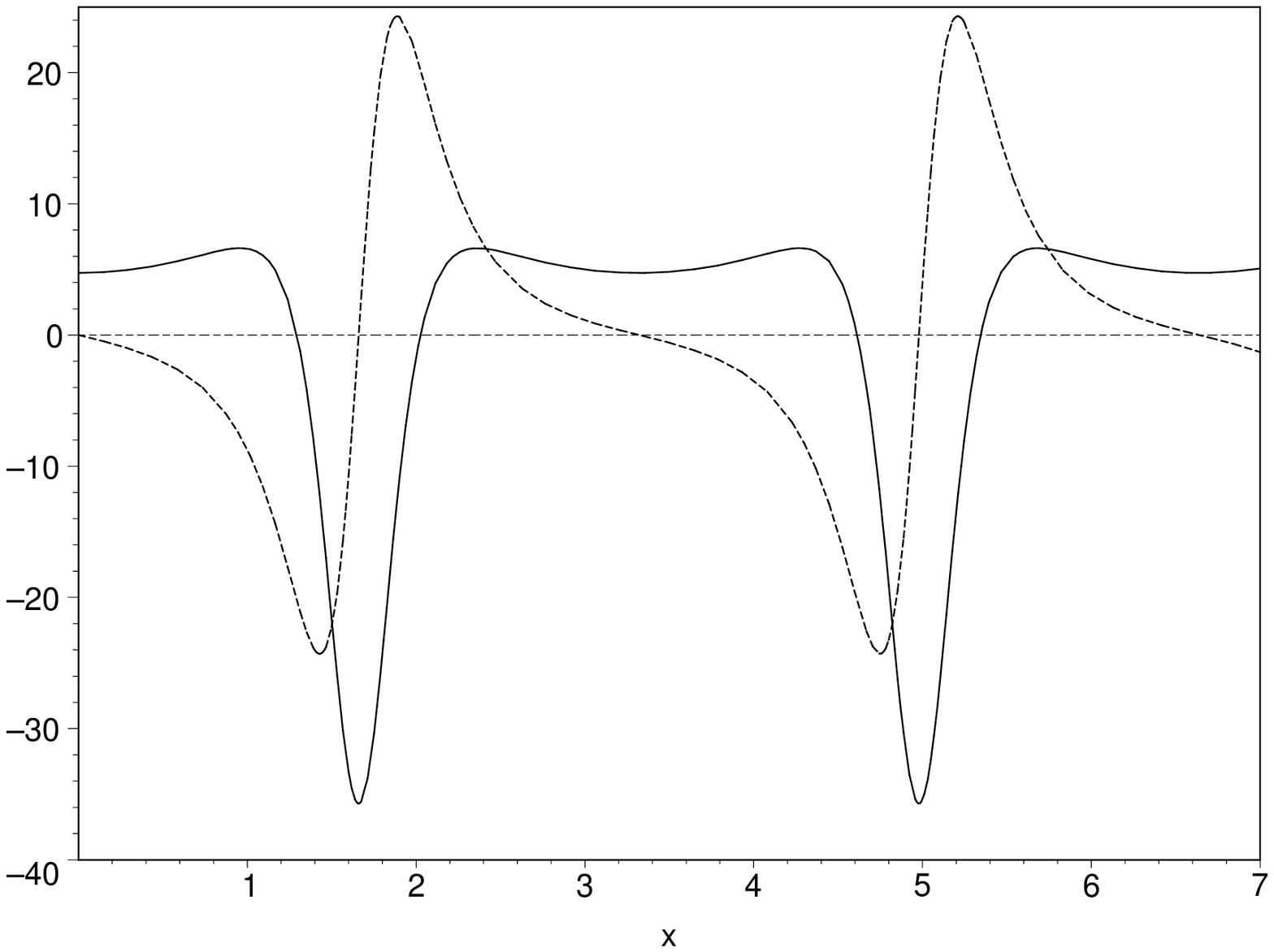,clip=,width=6.2in}
\caption{A plot of the real [solid line] and imaginary [dashed line] parts of the PT-invariant complex potential
(\ref{6.1}) for $a=2$. More explicitly, the potential is $-6m\sn^2 (ix+\beta,m)+2+2m+2 \sqrt{1-m+m^2}$ and has its lowest band edge at zero energy. The plot is made for the choice $m=0.8$ and $\beta =0.4$. The potential has a period $2K'(0.8)=3.3192$.   }
\end{figure}

\newpage


\begin{thebibliography}{99}
\bibitem{mw} W. Magnus and S. Winkler, {\it Hill's Equation} 
(Wiley, New York, 1966).
\bibitem{ar} F. M. Arscott, {\it Periodic Differential Equations} (Pergamon, 
Oxford, 1981). 
\bibitem{ww} E. T. Whittaker and G. N. Watson, {\it A Course of Modern 
Analysis} (Cambridge Univ. Press, Cambridge, 1980).
\bibitem{gr} For the properties of Jacobi elliptic functions, see, for example, 
I. S. Gradshteyn and I. M. Ryzhik, 
{\it Table of Integrals, Series and Products} 
(  Academic Press, 20000); 
M. Abramowitz and I. Stegun,
{\it Handbook of Mathematical Functions} (Dover, 1964).
The modulus parameter $m$ is often called $k^2$ in the 
mathematics literature. The related complementary quantity $(1-m)$ is often 
called $k'^2$.
\bibitem {ks1}  A. Khare and U. Sukhatme, Jour. Math. Phys. {\bf 40}, 5473 (1999).
\bibitem {ks2}  A. Khare and U. Sukhatme, Jour. Math. Phys. {\bf 42},
5652 (2001).
\bibitem{bed} C.M. Bender and S. Boettcher, Phys. Rev. Lett. {\bf 80}, 5243
(1998), For a recent review of this field, see C.M. Bender, D.C. Brody
and H.F. Jones, Amer. J. Phys. {\bf 71}, 1095 (2003) and references
therein.
\bibitem {ks3}  A. Khare and U. Sukhatme, math-ph/0402006. 
\bibitem{lau} See for example, D.F. Lawden, {\it Elliptic Functions
and Applications}, Applied Math. Sc. Vol. 80 (Springer, 1989); A.
Cayley, {\it An Elementary Treatise on Elliptic Functions} (G. Bell,
1895).  
\bibitem {ks4}  A. Khare and U. Sukhatme, math-ph/0208004. 
\bibitem {ks5}  A. Khare and U. Sukhatme, Jour. Math. Phys. {\bf 43},
5652 (2001); A. Khare, A. Lakshminarayan  and U. Sukhatme, ibid {\bf
44}, 1841 (2003); math-ph/0306028. 
\bibitem{cks} See, for example, F. Cooper, A. Khare and U. P. Sukhatme, Phys. 
Rep. {\bf 251}, 267 (1995).
\bibitem{df} G. Dunne and J. Feinberg, Phys. Rev. {\bf D57}, 1271
(1998).
\bibitem{bg} Y. Brihaye and M. Godart, J. Math. Phys. {\bf 34}, 5283 (1993); 
Y. Brihaye and S. Braibant, J. Math. Phys. {\bf 34}, 2107 (1993);
H. Braden and A. Macfarlane, J. Phys. {\bf A18}, 3151 (1995);
G. Dunne and J. Mannix, Phys. Lett. {\bf B428}, 115 (1998).
\bibitem{fer} D.J. Fern\'andez, B. Mielnik, O. Rosas-Ortiz and B.F.
Samsonov, Phys. Lett. {\bf A294}, 168 (2002); J. Phys. {\bf A35},
4279 (2002). 
\bibitem{kra} A. Krajewska, A. Ushveridze and Z. Walczak, Mod. Phys.
Lett. {\bf A12}, 1225 (1997).
\bibitem{bdm} C.M. Bender, G.V. Dunne and P.N. Meisinger, Phys. Lett. 
{\bf A252}, 272 (1999).  
\bibitem{raz} M. Razavy, Amer. J. Phys. {\bf 48},  285 (1980).
\bibitem{km} A. Khare and B.P. Mandal, J. Math. Phys. {\bf 39}, 3476
(1998).
\end{thebibliography}
\end{document}